# The influence of Y content on grain structure evolution in Mg-Y alloys

Qianying Shi[1], Vaidehi Menon[1], Liang Qi[1], John Allison[1,2]

[1]Department of Materials Science and Engineering, University of Michigan, Ann Arbor, MI 48109

[2]Department of Mechanical Engineering, University of Michigan, Ann Arbor, MI 48109

**Abstract:** To advance the understanding of microstructural evolution behavior in Mg-rare earth alloys, the effect of yttrium (Y) addition on static recrystallization and grain growth in Mg alloys was systematically investigated in extruded Mg-1wt.%Y and Mg-7wt.%Y alloys. Y addition was found to significantly retard the microstructural evolution, primarily due to its solute drag effect arising from Y segregation at grain boundaries. The relative intensity of solute drag effects from different alloying elements in Mg alloys was further assessed from both thermodynamic and kinetic perspectives, considering their grain boundary segregation tendencies and diffusivities. Additionally, static recrystallization in Mg-Y alloys was observed to proceed via a two-stage behavior characterized with two distinct JMAK exponents, indicating the heterogeneous nucleation of recrystallized grains. Abnormal grain growth (AGG) behavior was observed in these Mg-Y alloys. Overall, this study highlights the critical role of Y segregation at grain boundaries in controlling recrystallization and grain growth kinetics in Mg-Y alloys. This provides new insights into the design of thermally stable Mg alloys with refined microstructures.

# 1. Introduction

Magnesium-rare earth (Mg-RE) alloys have attracted significant attention since the 1950s due to their excellent mechanical performance and broad potential for lightweight structural applications [1]. The property enhancement in Mg-RE alloys largely benefits from RE alloying additions through two primary mechanisms: (i) extraordinary precipitation strengthening effect via the formation of thermally stable precipitates [2,3], and (ii) texture modification, commonly referred to as the "RE texture," which weakens the strong basal texture typically found in wrought Mg alloys and therefore enhances the formability of Mg-RE alloys [4–7].

Among various RE elements, yttrium (Y) has emerged as one of the most effective alloying additions to improve alloy performance. Y addition has been shown to improve ductility by increasing non-basal <c+a> slip activity and promoting contraction and double twinning [8–10]. Furthermore, Y reduces the $I_1$ stacking fault energy (SFE), therefore altering the relative critical resolved shear stresses for <a> and <c+a> dislocations [11–13]. In addition, Y exhibits a high solubility in Mg (up to ~11 wt.% at 572 °C [14]), which provides a strong driving force for solute segregation to grain boundaries and the activation of solute drag effects [15]. These unique effects of alloying element Y make the Mg-Y-based alloy system attractive for fundamental studies on deformation mechanisms and grain boundary physics, as well as for the design of advanced Mg alloys with superior mechanical performance.

As a key process in microstructural evolution, recrystallization plays an important role in restoring ductility, refining texture, and controlling grain size in Mg alloys. Extensive studies on the recrystallization behavior (both dynamic and static recrystallization) in conventional Mg alloys systems such as the AZ-series (Mg-Al-Zn-based) have been conducted and indicate that recrystallization is strongly influenced by solute content, deformation conditions, and annealing temperatures [16–36]. In contrast, for Mg-RE alloys, the majority of studies have focused on phase transformations (mainly precipitation behavior), texture evolution during thermomechanical process, deformation mechanisms, as well as mechanical and biological properties [37–44]. The recrystallization behavior for Mg-RE alloys (particularly Mg-Y alloys) remains less well understood, especially regarding the interaction between solute segregation and recrystallization nucleation/growth kinetics. Limited studies suggest that RE additions can retard recrystallization kinetics by reducing grain boundary mobility through solute drag effects, while also promoting heterogeneous nucleation via stored energy variations and segregation-induced boundary modifications [15,20,23,25,31,45]. These influences introduce complexity to the recrystallization process that has not been comprehensively characterized.

In addition to recrystallization, the grain growth behavior of Mg-RE alloys during annealing has also not been adequately studied compared to grain growth investigation for conventional Mg-Al-Zn-based alloys [33,46–51]. Recent reports indicated that abnormal grain growth (AGG) and associated texture changes may occur in certain Mg-RE alloys [52–54], highlighting the importance of elemental distribution at grain boundaries in these alloy systems. The occurrence of AGG is of particular interest because it implies strong heterogeneity in grain boundary mobility, which cannot be solely explained by traditional grain growth models. While mechanisms such as Zener pinning from second-phase particles or solute drag from grain boundary segregation have been widely considered [51,52,55], emerging evidence suggests that additional factors, such as anisotropic grain boundary mobility, grain boundary complexion transitions, and texture effects [53,56], may play critical roles in controlling AGG in Mg–RE alloys.

In this study, the static recrystallization and grain growth behaviors of two extruded Mg-Y alloys were systematically investigated during isothermal annealing heat treatment at different temperatures. Special emphasis was placed on clarifying the role of Y segregation at grain boundaries in governing recrystallization kinetics, grain boundary mobility, and AGG occurrence. Experimental observations are supported by solute drag models that incorporate rigorous calculations of solute segregation thermodynamics and existing literature on solute diffusion kinetics to provide new insights into the mechanisms of microstructural evolution in Mg-Y alloys. This study will contribute to the broader design of thermally stable Mg alloys with refined and controlled grain structures.

## 2. Materials and experimental procedure

Two extruded Mg-Y alloys with the nominal composition, Mg-1wt.%Y and Mg-7wt.%Y, were provided by CanmetMATERIALS, Natural Resources Canada. The alloy composition measured by inductively coupled plasma spectrometer (ICP) and the extrusion procedure were described in detail in a previous publication [57]. The Mg-1Y alloy was determined to have 0.93 wt.% Y and the Mg-7Y alloy was found to have 6.94 wt.% Y and 0.06 wt.% Ca.

For the recrystallization behavior study, cylindrical specimens (10 mm diameter and 12 mm height) were machined from the as-received extrusion bars using electrical discharge machining (EDM). The longitudinal axis of the specimens was aligned with the extrusion direction. Compression tests with a total strain of 20% along the longitudinal direction were performed using an Instron load frame equipped with a 100 kN load cell at a constant displacement rate of 0.03 mm/min. Because plastic strain was not uniformly distributed across the specimens [32], the central region was selected for microstructural characterization. Previous finite element analysis of this geometry determined that the plastic strain in this region was approximately 30-32% [32]. The compressed samples of two alloys were then annealed at 350 °C and 400 °C for varying times in a Thermolyne box furnace. For the grain growth study, as-extruded bars of each alloy were cut into ~10-15mm long sections and then subjected to annealing heat treatments at different temperature from 400 °C to 560 °C in the Thermolyne box furnace.

Standard grinding and polishing practices for Mg alloys were applied to prepare metallurgical samples. Buehler MetaDi fluid was used as the polishing lubricant instead of water, and the final polishing step employed 1 μm diamond paste. To reveal the microstructure in electron backscatter diffraction scanning (EBSD), polished specimens were etched in an acetic-nitric solution (60 mL ethanol, 20 mL water, 15 mL glacial acetic acid, and 5 mL of nitric acid) for ~15 seconds. EBSD scans were performed using a Tescan Mira3 electron microscope at 30 kV with a working distance of approximately 20 mm. The collected EBSD raw data were analyzed using either OIM software or MTEX, which is a free and open-source Matlab toolbox for analyzing and modeling crystallographic orientations [58].

To investigate the grain boundary segregation of solute elements, the focused ion beam (FIB) lift-out technique on a Thermo Fisher Helios 650 electron microscope was used to prepare site-specific foils for scanning transmission electronic microscope (STEM) examination. The STEM and energy dispersive X-Ray spectroscopy (EDS) analyses were performed on a Thermo Fisher Talos F200X TEM equipment operated at 200kV with a Super-X Quad windowless EDS detector.

3. Results

3.1. Initial microstructure

The initial grain structures of Mg-1Y and Mg-7Y alloys in the as-extruded condition are shown in **Figure 1**. Grains in both alloys were nearly fully recrystallized after the extrusion process. Differences in grain crystallographic orientations from the extrusion and radial directions were evident when comparing the EBSD inverse pole figure (IPF) maps, as shown in **Figure 1** (a1) and (a2) as well as **Figure 1** (b1) and (b2). These observations indicate a weak basal fiber texture, *i.e.* <0001>-*c* axis tends to weakly align with the radial direction and perpendicular to the extrusion direction (ED), while <2$\bar{1}\bar{1}$0> oriented grains tend to be aligned parallel to the ED (**Figure 1** (a3) and (b3) ).

The grain size distribution in the initial microstructures for the two alloys were similar, with an average grain diameter approximately 50 μm to 70 μm, although slight differences were noticed. The normalized grain size $D/\bar{D}$ was calculated for grains in **Figure 1** (a1) and (b1). The normalization parameter ($\bar{D}$) was defined as the area average grain diameter for each alloy. Grains at the edge of the image were excluded from this analysis. The distribution of $D/\bar{D}$ regarding to the grain area fraction for two alloys was then plotted as **Figure 1** (a4) and (b4), respectively. A Gaussian distribution for $D/\bar{D}$ was observed with the maximum $D/\bar{D}$ as 2.15 for Mg-1Y alloy and 1.86 for Mg-7Y alloy.

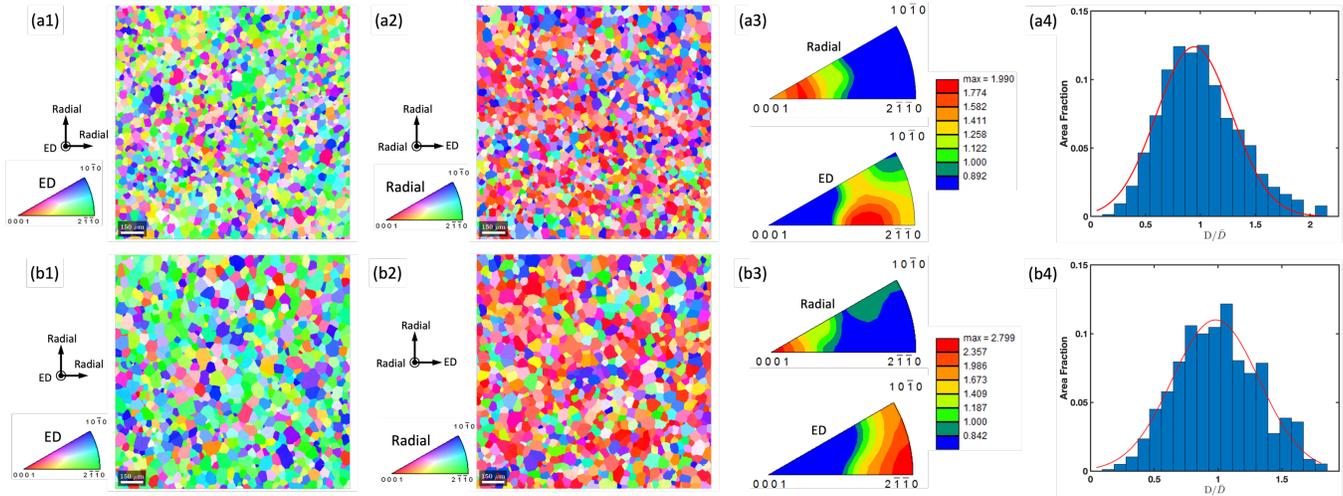

**Figure 1. Grain structure of Mg-1Y and Mg-7Y alloys in the as-received extruded bars shown in EBSD inverse pole figure (IPF) maps. (a) Mg-1Y; (b) Mg-7Y; (a1) and (b1) cross-sectioned, viewed from extrusion direction; (a2) and (b2) longitudinal, viewed from radial direction; (a3) and (b3) inverse pole figure texture analysis; (a4) and (b4) grain size distribution. Color code in IPF maps represents the crystallographic direction.**

**3.2. Static recrystallization behavior**

After the compression deformation with a total strain 20% at room temperature, both Mg-1Y and Mg-7Y alloys exhibited localized microstructural variations, although the overall microstructure appears homogeneous, as shown in **Figure 2** (a1) and **Figure 3** (a1). The resulting texture changes differed between two alloys. In Mg-1Y alloy, a strong deformation texture developed (**Figure 2** (a3)), with the <0001>-*c* axis tilted toward the extrusion direction, representing an ~90° shift of the *c*-axis compared with the initial texture (**Figure 1** (a3)). Mg-7Y alloy displayed similar deformation texture, however, the extent of the texture change before (**Figure 1** (b3)) and after deformation (**Figure 3** (a3)) was significantly lower than that observed in Mg-1Y alloy.

Grain orientation spread (GOS) maps from EBSD characterization, shown in **Figure 2** and **Figure 3**, were used to evaluate the degree of recrystallization before and after different annealing heat treatments. A GOS value of 1° was used as the cutoff between deformed and recrystallized grains. As seen in GOS maps in **Figure 2** (a2) and **Figure 3** (a2), both Mg-1Y and Mg-7Y alloys were in a completely deformed and unrecrystallized state.

For Mg-1Y alloy, annealing at 350°C for 2 hours resulted in almost complete recrystallization of the deformed grains, as shown in **Figure 2** (b1) and (b2). When the annealing temperature was increased to 400°C, a fully recrystallized microstructure was achieved in a much shorter time (~20 minutes) (**Figure 2** (c1) and (c2)). In contrast, the higher Y content was found to slow the recrystallization kinetic. In Mg-7Y alloy, a fully (or nearly fully) recrystallized structure was only obtained after 48 hours at 350 °C or 4 hours at 400 °C (**Figure 3**).

The recrystallization kinetics of both Mg-1Y and Mg-7Y alloys are compared in **Figure 4**, where the percentage of recrystallized grains (RX%) is plotted as a function of annealing time at both temperatures. The slower recrystallization kinetic of Mg-7Y alloy compared with Mg-1Y is clearly evident.

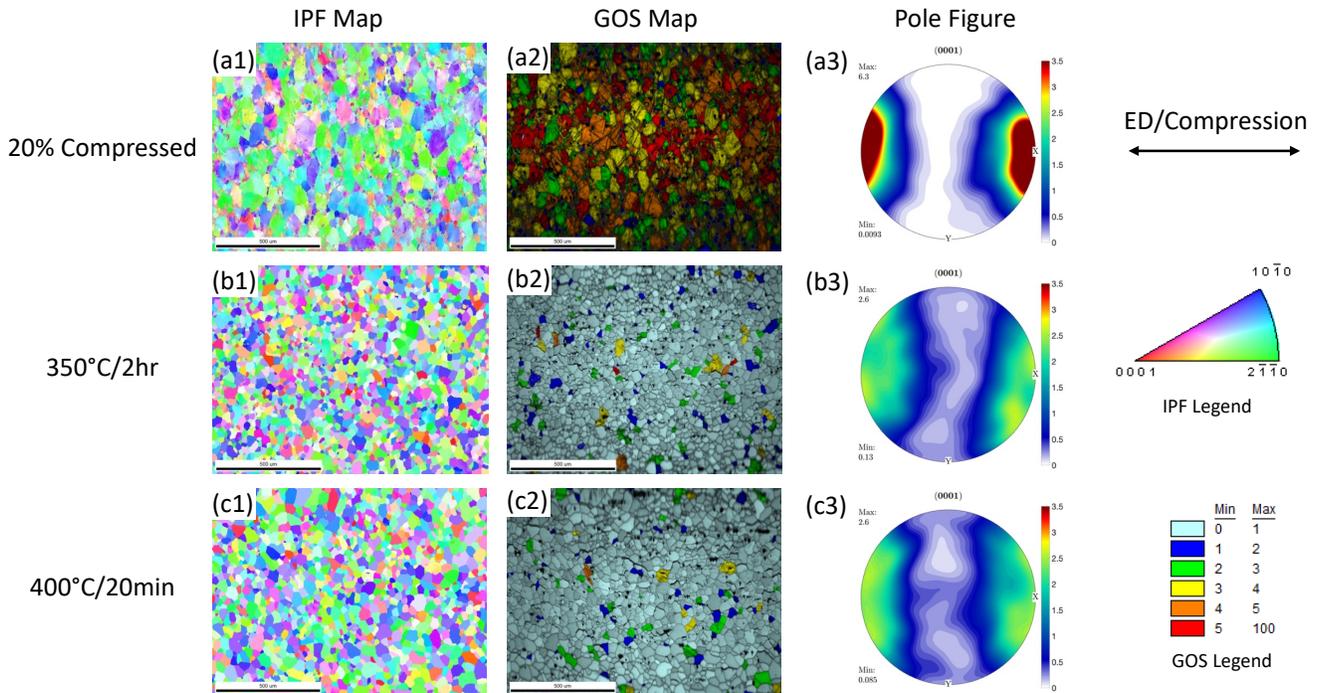

**Figure 2. Grain structure and texture of Mg-1Y alloy before and after the recrystallization heat treatment. (a1), (b1) and (c1) are IPF maps; (a2), (b2) and (c2) are GOS maps; (a3), (b3) and (c3) are pole figures; (a1)-(a3) are for Mg-1Y sample after 20% total strain compression; (b1)-(b3) are for Mg-1Y sample after the recrystallization heat treatment at 350°C for 2 hours; (c1)-(c3) are for Mg-1Y sample after the recrystallization heat treatment at 400°C for 20 minutes. The compression axis (*i.e.* extrusion direction) is horizontal. Color code in IPF maps represents the crystallographic direction of grains along the radial direction.**

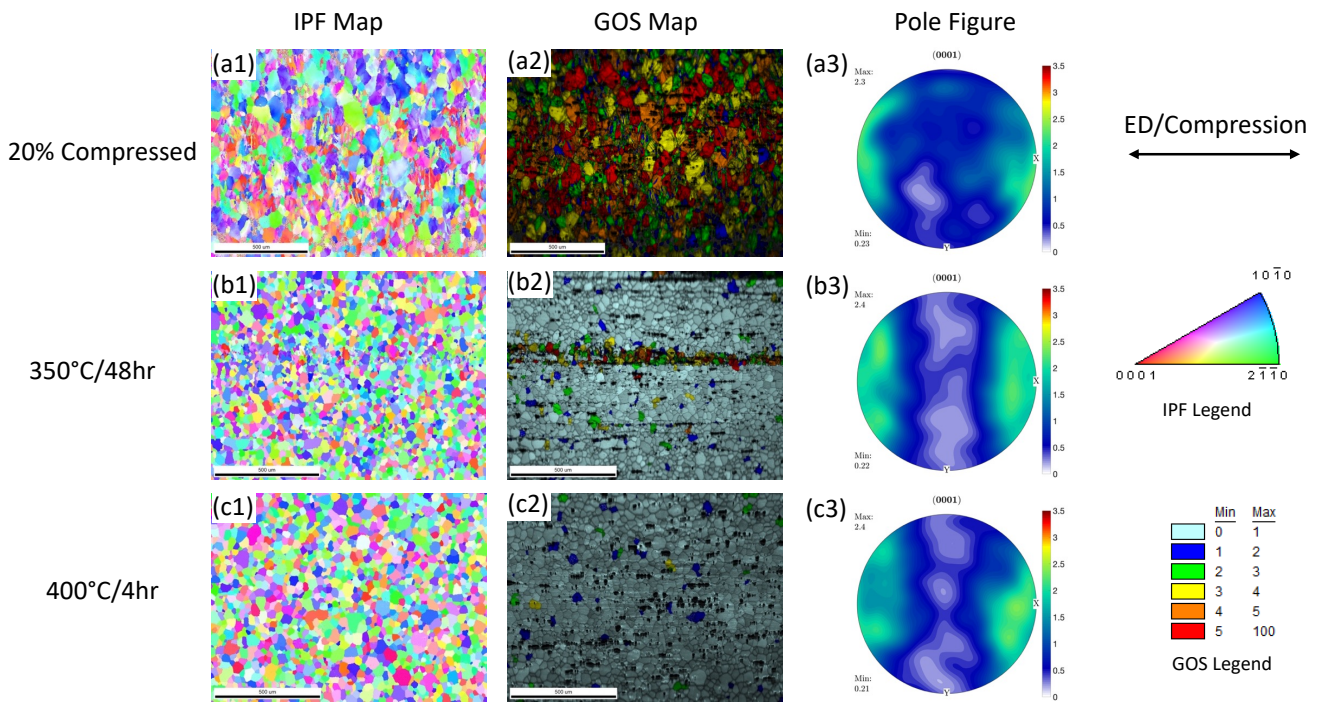

**Figure 3. Grain structure and texture of Mg-7Y alloy before and after the recrystallization heat treatment. (a1), (b1) and (c1) are IPF maps; (a2), (b2) and (c2) are GOS maps; (a3), (b3) and (c3) are pole figures; (a1)-(a3) are for Mg-7Y sample after 20% total strain compression; (b1)-(b3) are for Mg-7Y sample after the recrystallization heat treatment at 350°C for 48 hours; (c1)-(c3) are for Mg-7Y sample after the recrystallization heat treatment at 400°C for 4 hours. The compression axis (*i.e.* extrusion direction) is horizontal. Color code in IPF maps represents the crystallographic direction of grains along the radial direction.**

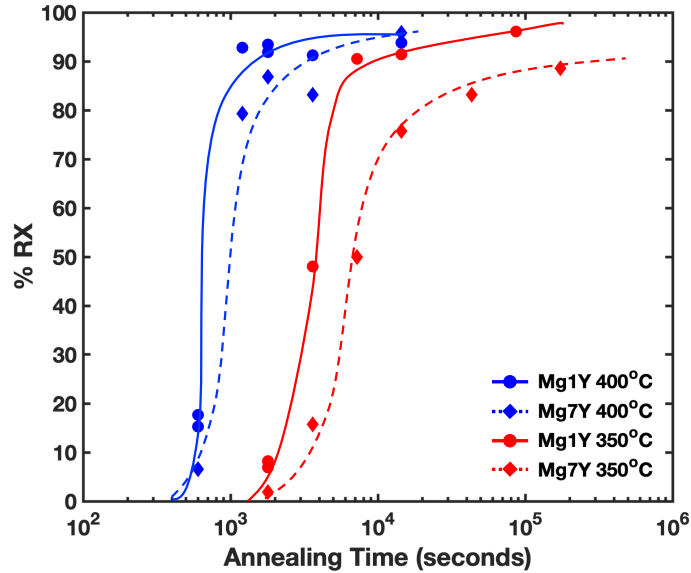

**Figure 4. Fraction of recrystallized grains as a function of annealing time at 350°C and 400°C for both Mg-1Y and Mg-7Y alloys. The trendline for each condition was added for visual assistance.**

### 3.3. Grain growth behavior

The grain growth behavior of the two extruded Mg-Y alloys was studied at a wide range of temperatures: 400°C, 450°C, 500°C, 520°C, 540°C, and 560°C. For Mg-1Y alloy, no grain growth was observed at 400°C, even after long annealing times (up to 96 hours). At 450°C, grain growth was not evident after 8 hours (**Figure 5** (a)) but became apparent after 12 hours ((**Figure 5** (b)). Interestingly, the microstructure at 450°C for 12 hours showed a bimodal grain size distribution, with both very large grains (grain diameters up to ~1000 μm) and small grains (comparable to the initial grain size) coexisting. At higher temperatures (≥500°C), grains in Mg-1Y alloy grew rapidly, making the early stages of grain growth difficult to capture. For example, after annealing at 520°C for 15 minutes, the average grain size had already increased to ~300 μm (**Figure 5** (c)), which is approximately 5-6 times larger compared to the initial microstructure.

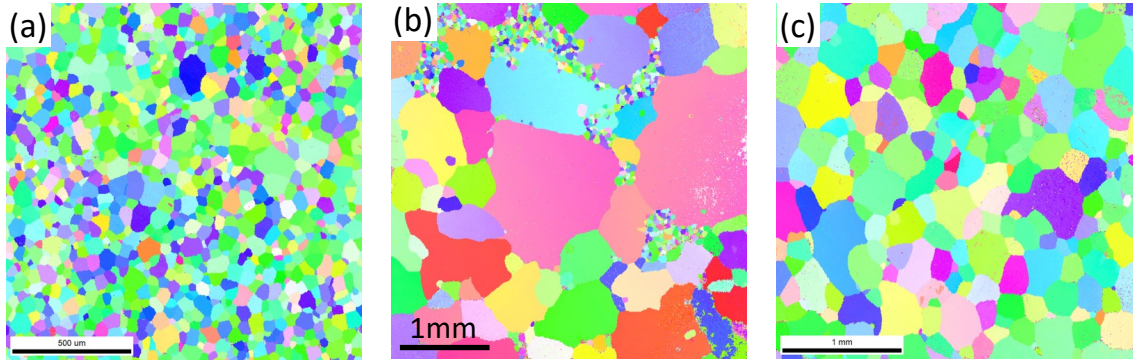

**Figure 5. EBSD-IPF maps showing the grain structure and corresponding grain size distribution for Mg-1Y alloy at different annealing conditions. (a) 450°C for 8 hours; (b) 450°C for 12 hours; (c) 520°C for 15 minutes. Color code in IPF maps represents the crystallographic direction with reference to the extrusion direction.**

**Figure 6** shows the grain size as a function of annealing time at different annealing temperatures for Mg-7Y alloy. In this high Y containing alloy, grain growth occurred more slowly than in Mg-1Y alloy. No significant grain growth was observed at the relatively low annealing temperature, 400°C, 450°C and 500°C with annealing time up to 24 hours. As the temperature was elevated to 520°C, a slight increase in grain size was detected after 48 hours ($1.7 \times 10^5$ seconds). As further increasing the temperature to 540°C, an obvious grain growth was observed after 8 hours ($2.9 \times 10^4$ seconds). At the highest annealing temperature studied (560°C), a sharp increase in the average grain size occurred after only 30 minutes of annealing. Grains continued to grow during prolonged annealing at 560°C. The average grain size reached approximately 600 μm after 24 hours, which is nearly an order of magnitude larger than the initial grain size.

It is worth noting that the standard deviation of grain size was large under annealing conditions where grain growth occurred, indicating a broad grain size distribution. **Figure 7** presents EBSD-IPF maps of Mg-7Y alloy at selected annealing conditions where grain growth was evident. Abnormally large grains (ALGs), which is defined as grains significantly larger than the surrounding ones, were observed at 540°C after 8hours (**Figure 7** (a1)), 540°C after 24hours (**Figure 7** (b1)) and 560°C after 30minutes (**Figure 7** (c1)). Due to the appearance of ALGs, the normalized grain size distributions broadened with maximum D/$\overline{D}$ values approaching ~3, and deviated from Gaussian fits, as shown in **Figure 7** (a2), (b2) and (c2).

When annealing at 560°C was extended to 1 hour, most grains had grown (**Figure 7** (d1)), and the range of normalized grain size distribution became narrow with the maximum $D/\overline{D}$ as 1.97 (**Figure 7** (d2)). After 24 hours at 560°C, grains continued to grow, and the D/$\overline{D}$ distribution remained narrow (**Figure 7** (e2)). However, the distribution deviated from a Gaussian shape. This is likely due to the limited number of grains (107) contributing to the plot, as grain sizes became very large under this condition.

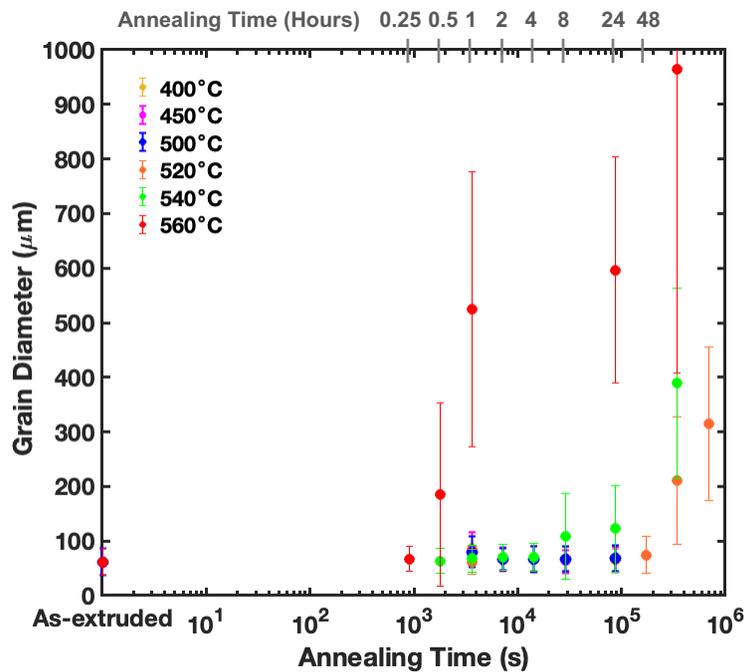

**Figure 6. Grain diameter as a function of annealing time for Mg-7Y alloy at annealing temperatures ranging from 400°C to 560°C.**

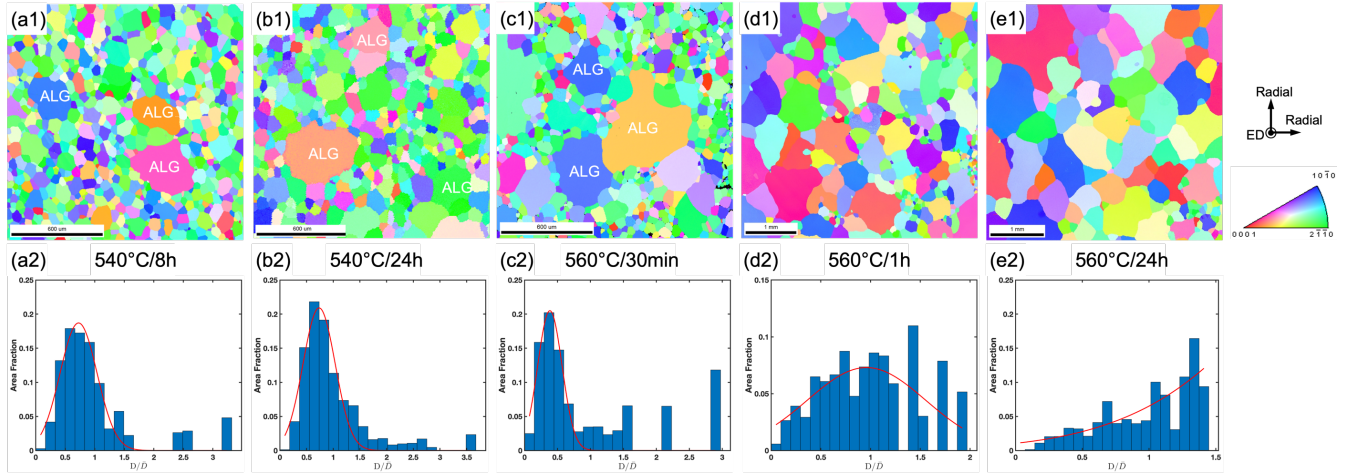

Figure 7. EBSD-IPF maps showing the grain structure and corresponding grain size distribution of Mg-7Y for different annealing conditions. (a1) and (a2) 540°C/8h; (b1) and (b2) 540°C/24h; (c1) and (c2) 560°C/30min; (d1) and (d2) 560°C/1h; (e1) and (e2) 560°C/24h. Color code in IPF maps represents the crystallographic direction with reference to extrusion direction. Red lines in (a2), (b2), (c2), (d2), (e2) are Gaussian function fitting curves.

### 3.4. Crystallographic orientation analysis on ALGs

Comparison of EBSD-IPF maps between the initial microstructure (**Figure 1** (b1)) and those after significant grain growth (**Figure 7** (d1) and (e1)) reveals that the texture of Mg-7Y alloy changed as grains grew. To evaluate this crystallographic orientation evolution during grain growth, a large area (4500 μm by 4500 μm) of the sample annealed at 560°C for 30 minutes was scanned for EBSD data collection to include a representative number of grown grains. The EBSD analysis results are shown in **Figure 8**.

Based on the grain size distribution (**Figure 8** (b)), grains were separated into two groups using an arbitrary cutoff diameter of 180 μm, approximately three times the average grain diameter in the as-extruded condition. Grains smaller than 180 μm were considered as non-growing small grains, since they are still in the range of the initial grain size distribution. Grains with diameters larger than 180 μm were classified as abnormally large grains. As shown in **Figure 8**(c), the IPF map of small grain group was dominated by green coloring, consistent with the initial as-extruded texture ($<2\bar{1}\bar{1}0>$ parallel to the extrusion direction). In contrast, the large-grain group (**Figure 8** (d)) exhibited more blue and purple coloring, with the maximum texture intensity shifted to a new orientation: $<10\bar{1}0>$ parallel to the extrusion direction.

This texture evolution during grain growth was further confirmed in the sample annealed at 560°C for 24 hours. **Figure 9** shows the EBSD analysis of this condition. After 24 hours, nearly all grains had grown to varying degrees, as indicated by the grain size distribution (**Figure 9** (a)), compared to the initial state. The inverse pole figure analysis (**Figure 9** (b) and (c)) confirmed the development of a $<10\bar{1}0>$//ED texture in the grown grains, suggesting that grains with this orientation had a growth.

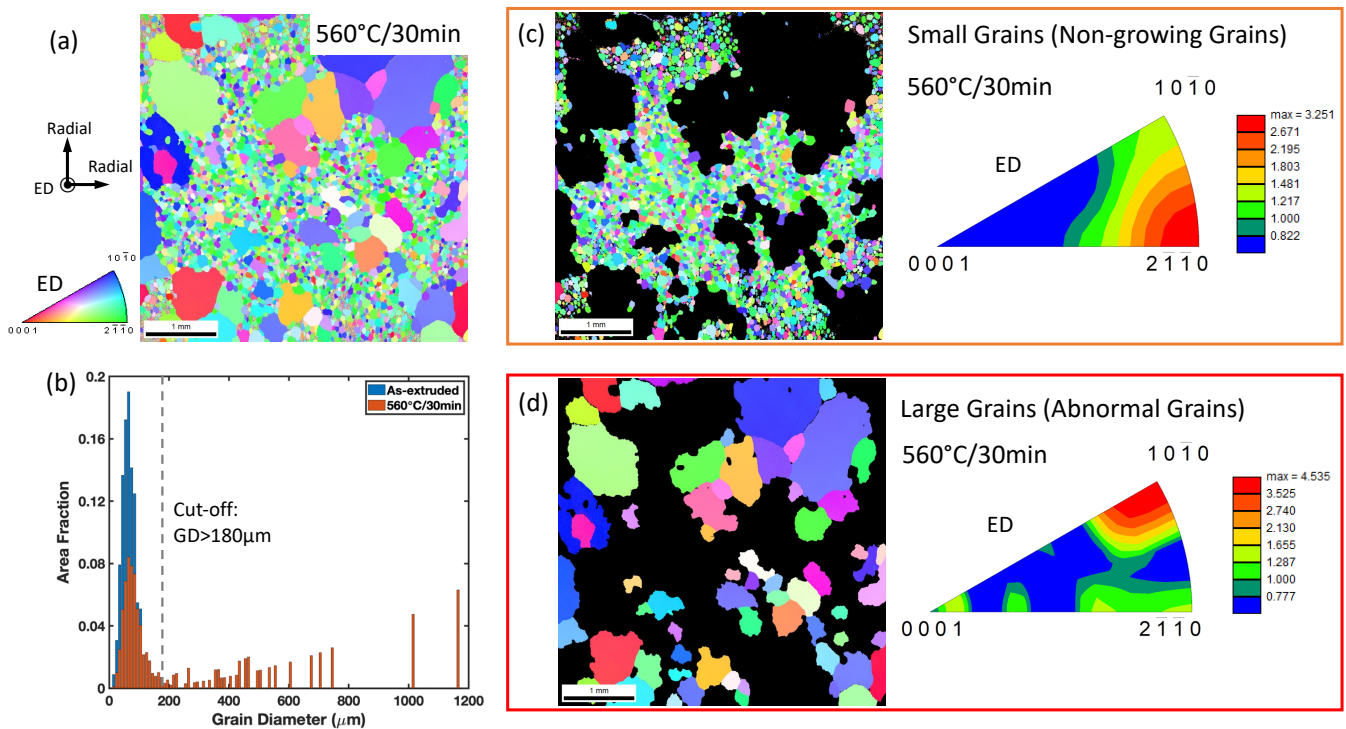

**Figure 8. Texture differences after grain growth in Mg-7Y alloy. (a) EBSD-IPF map for a large area to include a representative number of large abnormal grains for the annealing condition at 560°C for 30minutes; (b) Grain diameter distribution plots for the as-extruded condition and the annealed condition at 560°C for 30 minutes. (c) and (d) show the texture of small, non-growing grains (c) and large abnormal grains (d). All inverse pole figure texture analysis and grain maps are with respect to the extrusion direction.**

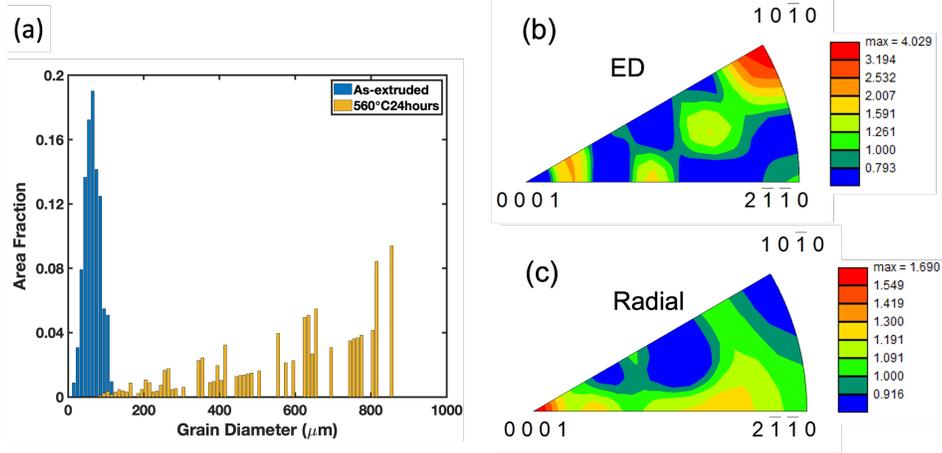

**Figure 9.** EBSD data analysis for the annealed sample at 560°C for 24 hours. (a) Grain diameter distribution compared with the as-extruded condition; (b) and (c) Inverse pole figures viewed from extrusion direction and radial direction, respectively.

## 4. Discussion

### 4.1. Y effect on static recrystallization

The recrystallization kinetics in **Figure 4** exhibited the traditional sigmoidal-shaped curves, which signals that the well-known Johnson-Mehl-Avrami-Kolmogorov (JMAK) equation can be used to quantify this process. The JMAK relationship in equation [1] relates the fraction of recrystallized grains (X) to the annealing time (t), in which $t_{0.5}$ is the time required to achieve 50% recrystallization and $n$ is the Avrami exponent. In order to calculate $t_{0.5}$ and $n$ values, values of ln(ln(1/(1-X))) and ln(t) were calculated using the experimental data, and then plotted versus each other, as shown in **Figure 10**. The slope of the linear fitting line for each condition was measured, which is the $n$ value, while $t_{0.5}$ was then calculated from the y-intercept. According to equation [2], the activation energy for recrystallization $Q_{rex}$ and pre-exponential factor $A\epsilon^{-p}$ can be determined from the slope and y-intercept of the linear fit for $\ln(t_{0.5})$ and 1/RT.

$$X = 1 - exp\left(-0.693\left(\frac{t}{t_{0.5}}\right)^n\right) \qquad [1]$$

$$t_{0.5} = A\epsilon^{-p} exp\left(\frac{Q_{rex}}{RT}\right) \qquad [2]$$

It is noticeable in **Figure 10** that a single linear fit was not able to capture the entire recrystallization process. It appeared to be bilinear, which is representative of a two-stage recrystallization process. Therefore, two different Avrami exponents, $n_1$ and $n_2$, were determined separately. These JMAK parameters were summarized in **Table 1** and compared with Mg-4wt.%Al alloy and pure Mg, which were reported in Reference [32]. These two stages of JMAK relationship were able to capture the static recrystallization kinetics for each condition in this study well. As shown in **Figure 11**, a rapid initial recrystallization occurred which was then followed by an abrupt retardation of recrystallization which occurred after approximately 75% to 90% recrystallization.

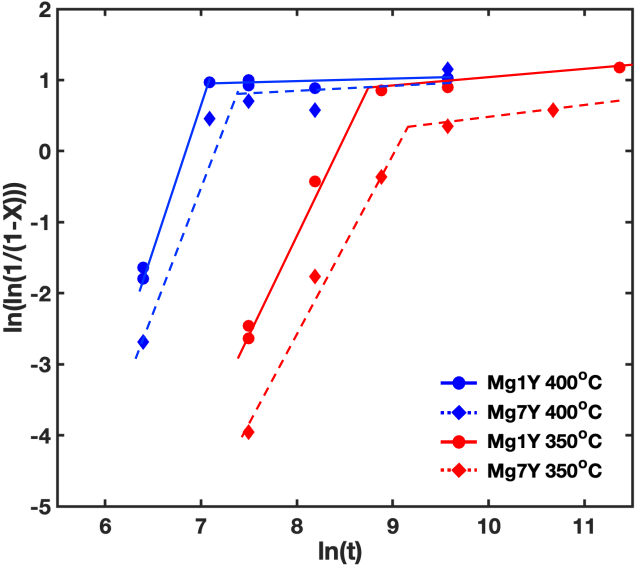

**Figure 10. Plots of ln(ln(1/(1-X))) vs ln(annealing time) used to determine the JMAK exponents at each annealing temperature for the two Mg-Y alloys.**

**Table 1.** JMAK constants for different alloys at each annealing temperature

| Alloy | Temperature (°C) | $n_1$ | $n_2$ | $t_{0.5}$ (s) | $Q_{rex}$ (kJ/mol) | $A\epsilon^{-p}$ |
|---|---|---|---|---|---|---|
| Mg-1Y | 350 | 2.51 | 0.13 | 4162 | 111 | $2.5 \times 10^{-6}$ |
| Mg-1Y | 400 | 3.87 | 0.02 | 850 | | |
| Mg-7Y | 350 | 2.59 | 0.17 | 6846 | 125 | $2.7 \times 10^{-7}$ |
| Mg-7Y | 400 | 3.24 | 0.25 | 1139 | | |
| Mg-4Al [32] | 250 | 0.45 | 0.08 | 7200 | | |
| Mg-4Al [32] | 275 | 0.51 | 0.20 | 2800 | 177 | $1.9 \times 10^{-14}$ |
| Mg-4Al [32] | 300 | 0.48 | 0.05 | 200 | | |
| Pure Mg [32] | 150 | 0.52 | 0.12 | 1000 | | |
| Pure Mg [32] | 200 | 0.59 | 0.02 | 300 | 123 | $1.1 \times 10^{-12}$ |
| Pure Mg [32] | 250 | 0.36 | 0.05 | 0.95 | | |

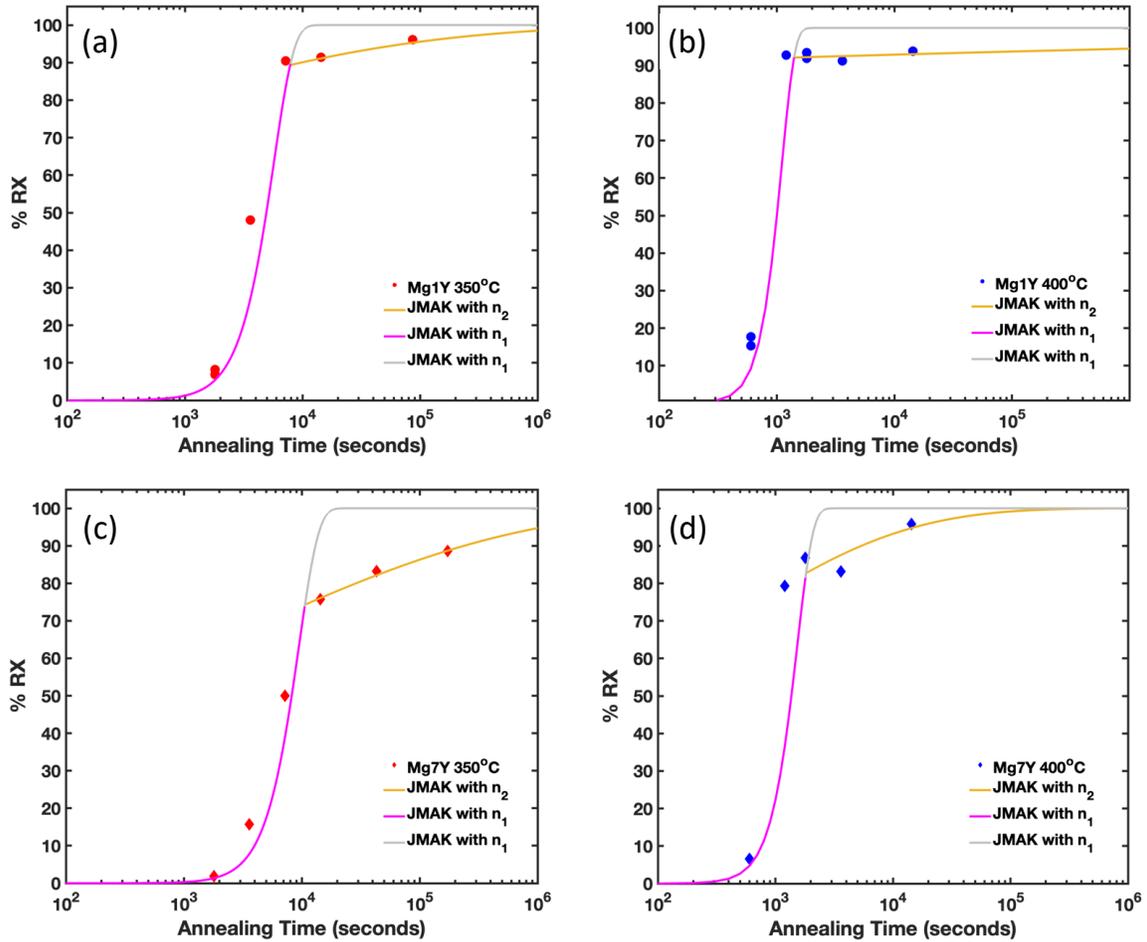

**Figure 11.** Two stages of JMAK relationship to capture the static recrystallization kinetics for each condition. (a) Mg-1Y alloy, 350°C; (b) Mg-1Y alloy, 400°C; (c) Mg-7Y alloy, 350°C; (d) Mg-7Y alloy, 400°C.

Static recrystallization in metallic systems with two distinct stages have also been reported in a range of different alloy systems, including Ti [59], steel [60,61] and Mg-alloys [16,32]. This behavior has generally been attributed to microstructural inhomogeneities and heterogeneous nucleation mechanisms. The initial, rapid stage is controlled by the preferential nucleation at sites with high stored energy, such as deformation-induced defects (shear bands), grain boundaries and twin boundaries, which serve as favorable locations for the formation of new grains due to their high defect densities [16,32,59–61]. Once these regions are fully recrystallized, the process enters a slower second stage, during which nucleation and growth occur within the interiors of grains that experienced comparatively lower strains. In the case of Mg alloys, twinning is a particularly common deformation mechanism that would significantly influence the recrystallization behavior. A previous study has reported the evidence of nucleation of recrystallized grains within twins [32], underscoring their importance as active sites for the onset of recrystallization. While twinning provides an important complementary pathway that facilitates the recrystallization process, grain boundaries are still recognized as the dominant nucleation sites for recrystallized grains [32]. Similar heterogeneous nucleation mechanisms for recrystallization can be anticipated in current investigated Y-containing Mg alloys. In related studies [62,63], we have determined that twinning is reduced in Mg-7Y compared to Mg-1Y and both alloys have substantially lower twin activity compared to pure Mg and Mg-Al alloys.

As shown in **Table 1**, the Stage 1 Avrami exponent $n_1$ for both Mg-1Y and Mg-7Y alloys is approximately three, which corresponds to the typical value for site-saturated nucleation [64]. An exponent of this magnitude indicates that the nucleation of new, strain-free grains effectively occurs at the start of recrystallization process, and the nucleation sites are randomly distributed. The $n_1$ values obtained in this study are significantly higher than those reported for pure Mg ($n_1$ ~0.5) [32], Mg-4Al ($n_1$ ~0.5) [32] and AZ31 ($n_1$ ~1) [33,34]. The substantially lower Avrami exponents in those alloys were attributed to preferred nucleation at twins and grain boundaries. Although the heterogeneous nucleation is expected in these Mg-alloys, the difference of Avrami exponent $n_1$ between Y-containing alloys and others suggests a comparatively lower degree of heterogeneity in nucleation in Mg-Y. This behavior can be rationalized by the weaker texture and reduced twin activity observed in Y-containing alloys [62,63], both of which decreased the availability of preferential nucleation sites such as twin boundaries. Consequently, the static recrystallization in Mg-Y alloys appeared to proceed through a relatively more spatially uniform nucleation process than that in other Mg-based alloy systems.

The activation energy for recrystallization ($Q_{rex}$) was determined to be 111 kJ/mol for Mg-1Y alloy and 125 kJ/mol for Mg-7Y alloy, as listed in **Table 1**. These values fall within the range reported for other Mg alloys. For instance, $Q_{rex}$ values of 177 kJ/mol and 123 kJ/mol have been reported for Mg-4Al and pure Mg, respectively [32], while values of 200 kJ/mol and 125 kJ/mol were determined for AZ31 alloys in the studies of Beer et al. [19,35] and Yang et al. [36]. For reference, the activation energy for self-diffusion of Mg has been established at approximately 135 kJ/mol [19]. These comparisons suggest that recrystallization in Mg and its alloys is predominantly governed by Mg self-diffusion. It should be noted, however, that the direct comparison of $Q_{rex}$ values across different studies is complicated due to variations in experimental methodologies.

In addition, the pre-exponential factors $A\epsilon^{-p}$ for Mg-Y alloys are substantially higher than those for pure Mg and Mg-4Al alloy (**Table 1**). This indicates that a significantly longer annealing time is required for Mg-Y alloys to achieve the same level of recrystallization as in other Mg alloy systems. The parameter $A$ reflects contributions from several kinetic factors, including the grain boundary mobility. Previous studies have shown that solute effects on the recrystallization behavior are primarily attributed to their influence on grain boundary mobility [36,65–67]. Our earlier work demonstrated that Y exhibits a strong tendency to segregate at grain boundaries [57], and higher Y additions can also promote deformation-induced segregation at twin boundaries [62]. To further examine this behavior, a grain boundary in a fully recrystallized Mg-7Y sample subjected to annealing at 400°C for 4 hours was analyzed using STEM-EDS. As shown in **Figure 12**, Y segregation was clearly revealed at the boundaries of recrystallized grains. This direct evidence strongly supports the conclusion that the solute drag from Y segregation contributes to lower the grain boundary mobility and thus sluggish recrystallization kinetics observed in Mg-Y alloys.

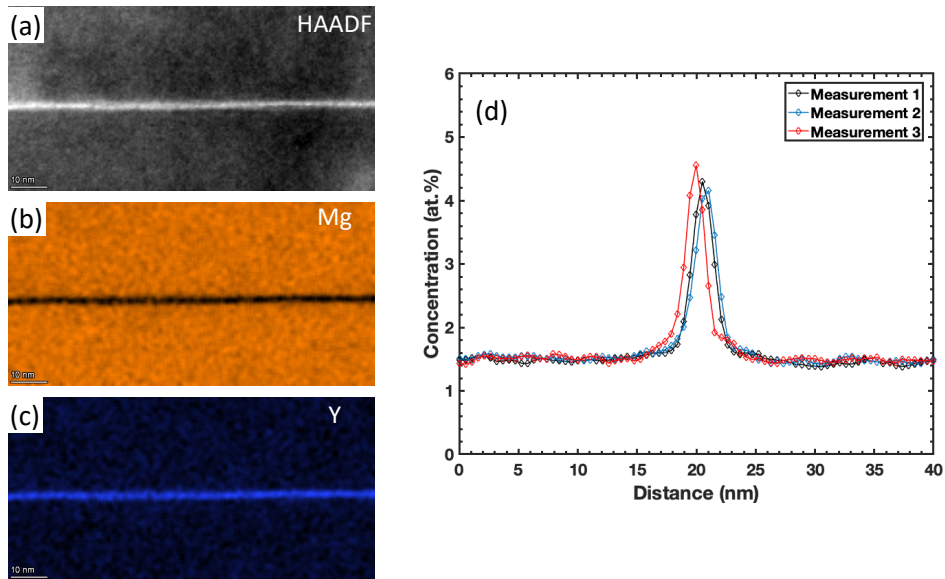

**Figure 12. Grain boundary segregation characterized by STEM and EDS on Mg-7Y alloy sample after a fully recrystallization heat treatment at 400°C for 4 hours. (a) HAADF image taken with the grain boundary tilted close to edge-on condition; (b) and (c) EDS mapping results for elements Mg and Y; (d) Integrated EDS line analysis of Y concentration across the grain boundary.**

### 4.2. Y effect on grain growth

As described in Section 3.3, the grain growth rate of Mg-7Y alloy was significantly lower than that of Mg-1Y. This enhanced microstructure stability at higher Y content alloy contrasts with the behavior of pure Mg and Mg-4wt.%Al alloy investigated in the previous study [68]. Those two materials were produced under comparable extrusion conditions, and their as-extruded microstructures were similar to that of Mg-Y alloys in the present study with the exception that the Mg and Mg-4Al exhibited a more pronounced basal texture. For a direct comparison, the grain growth curves of three alloys at comparable annealing temperatures (450°C/435°C and 520°C/515°C) are plotted in **Figure 13**(a) and (b), respectively. The results reveal that Mg-7Y alloy exhibits markedly higher resistance to the grain growth, as the considerable grain growth occurred in both pure Mg and Mg-4Al under these conditions, whereas grain growth in Mg-7Y remained strongly suppressed. This effect is most pronounced at the higher (520°C/515°C) annealing temperatures.

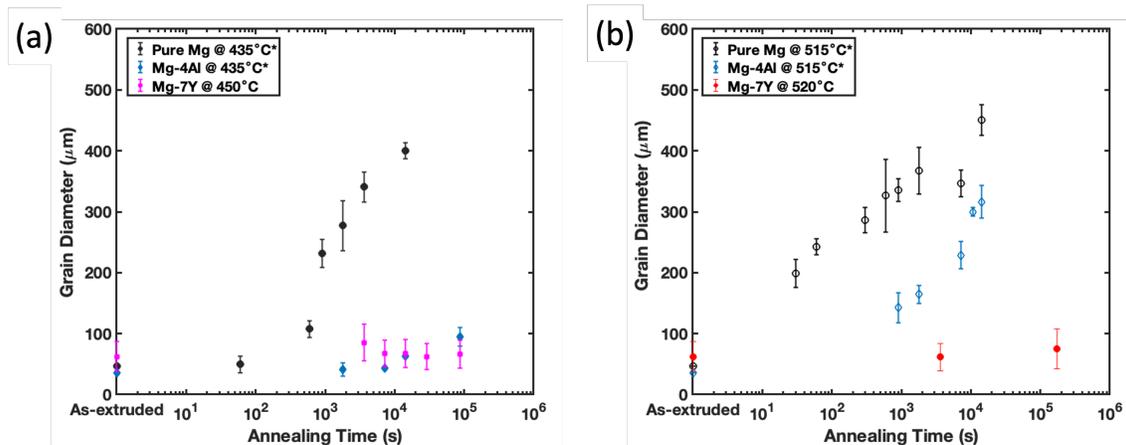

**Figure 13. A comparison of grain growth behavior inMg-7Y, pure Mg and Mg-4Al. (a) Relatively low annealing temperature 450°C/435°C; (b) Relatively high annealing temperature 520°C/515°C. *Experimental results from the Reference** [68]**.**

Grain growth, whether occurring in the normal grain growth (continuous) or abnormal (discontinuous) mode, is fundamentally controlled by grain boundary migration [69,70]. Consequently, any factor that impedes the boundary mobility will retard grain growth process. These factors typically include second-phase particles, solute elements, and grain boundary characteristics such as misorientation [70,71]. Among these, second-phase particles are widely recognized to apply a drag force on the grain boundaries, inhibiting the boundaries from moving, which is commonly referred to as the Zener pinning effect. For instance, Bhattacharyya et. al claimed that the presence of second-phase particles in different type and size (Al-Mn-based, Mg-Al-based and impurity-based particles) influenced the grain growth kinetics of AZ31B alloy [51]. Similarly, Alizadeh et al. demonstrated that $Mg_3(Gd,Y)$ particles in Mg-Gd-Y-Zr alloy restricted grain growth through particle pinning the grain boundaries [55]. In the present Mg-7Y alloy, a limited volume fraction (~0.15%) of Y-enriched particles was observed in the as-extruded microstructure. Notably, these particles were randomly distributed along the grain boundaries and in grain interiors. Although their effect is not considered dominant, they may still contribute to the reduced grain boundary mobility and, consequently, to the sluggish grain growth observed in Mg-Y alloys.

As with second-phase particles, solute elements segregated at grain boundaries are believed to produce a solute drag pressure on migrating boundaries. Our previous work demonstrated that Y exhibits a strong tendency to segregate at grain boundaries [57]. In the present alloys, Y enrichment at grain boundaries was experimentally observed both in the as-extruded condition (prior to grain growth) and after high-temperature annealing (following grain growth) [57]. Consequently, grain boundary motion in Mg-Y alloys, particularly in the Mg-7Y alloy, was significantly impeded by the solute drag effect associated with Y segregation. As a result, the alloy with higher Y content exhibited sluggish grain growth kinetics.

**4.3. Occurrence of abnormal grain growth**

**4.3.1 Quantification of abnormal grain growth**

Abnormal grain growth (AGG) refers to a coarsening type of microstructure in which a subset of grains grows unusually rapidly within a matrix of fine grains that exhibit very limited growth [72]. AGG has been widely reported in different materials including steels [73–76], titanium [77], aluminum [78], as well as magnesium and its alloys [46,51–53,56,79]. A common approach to quantitatively distinguish normal grain growth (NGG) from AGG is through the normalized grain size distribution $D/\overline{D}$. During NGG, a narrow fixed-form unimodal distribution remains essentially unchanged throughout the growth process. This normal state is typically defined such that the largest grain is smaller than twice that of the critical grain size, size which is determined as 9/8 times of the average grain size [70,80]. In contrast,

AGG is characterized by the loss of self-similarity in the grain size distribution, often reflected by the development of a bimodal $D/\bar{D}$ distribution or a significant broadening of the $D/\bar{D}$ distribution [81].

In the present Mg-Y alloys, the initial microstructure exhibited a relatively narrow grain size distribution, with the largest grain smaller than twice the critical grain size, as shown in **Figure 1** (a3) and (b3). Although no clear bimodal distribution was detected, the $D/\bar{D}$ distributions for Mg-7Y alloy at three annealing conditions, 520°C/48h, 540°C/8h and 560°C/30min, became noticeably broadened with extended tails, as shown in **Figure 7** (a2), (b2) and (c2). This suggests the occurrence of AGG, and the initial normal state was lost at these annealing conditions. In addition, it is interesting to note that AGG was transient. As the annealing time was increased to 1 hour at 560°C, the $D/\bar{D}$ distribution reverted to a narrow, nearly unimodal form, as shown in **Figure 7** (d2). It is indicated that the abnormally large grains experienced impingement and the fine grains in the matrix quickly caught up through NGG. Such transient abnormal grain growth and the subsequent transition to NGG have also been reported in earlier studies [51,82–84].

Another indicator proposed to identify AGG is the positive derivative of $D/\bar{D}$ [53,72]. To evaluate this in the present study, the ratio ($D_{max}/\bar{D}$) was analyzed where between the maximum grain diameter $D_{max}$ is the maximum grain diameter for a particular annealing condition and $\bar{D}$ is the average grain diameter of all grains. **Figure 14** shows the variation of $D_{max}/\bar{D}$ as a function of the annealing time at different temperatures for Mg-7Y alloy. Although there are some fluctuations in this value, it is clear that a significant increase of $D_{max}/\bar{D}$ occurs at 520°C/48h, 540°C/8h and 560°C/30min. This indicates a positive derivative of $D_{max}/\bar{D}$ and is an indication of the occurrence of AGG. As increasing the annealing time to 1 hour at 560°C, the $D_{max}/\bar{D}$ dropped significantly, *i.e.* indicating a negative derivative of $D_{max}/\bar{D}$ and a reversion to NGG. This further confirms the transient nature of AGG, with the grain growth mode transitioning back to NGG within a relatively short change in annealing time.

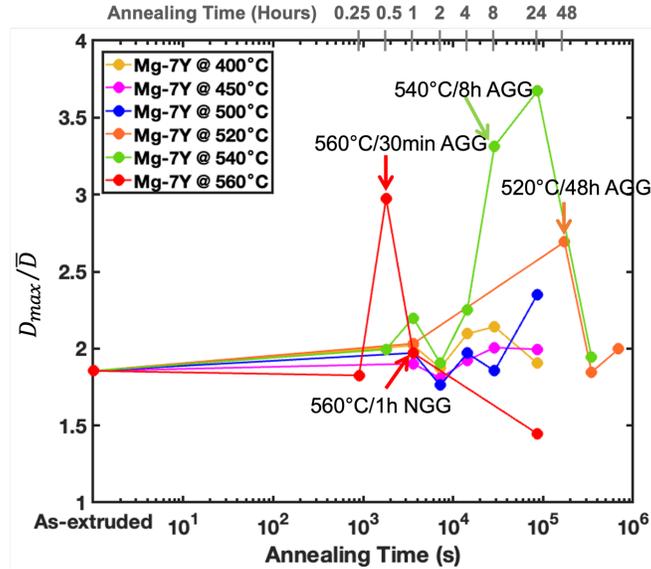

**Figure 14.** The indicator $D_{max}/\overline{D}$ of abnormal grain growth as a function of annealing time for different temperatures.

### 4.3.2 Mechanisms for abnormal grain growth

Several mechanisms have been proposed to explain the occurrence of AGG, although some remain controversial. Bhattacharyya et al. reported AGG in a strongly textured AZ31B Mg alloy under annealing conditions at 300°C to 450°C, attributing its occurrence primarily to Zener pinning by second-phase particles [51]. Basu et al. proposed a different mechanism in Mg-Gd alloys, suggesting that at sufficiently high annealing temperatures, certain grain boundaries could overcome the drag effect associated with Gd segregation and continue to grow, while others remained pinned [52]. However, AGG was also observed in hot-rolled, non-alloyed pure-Mg annealed at an intermediate temperature of 220°C, as reported by Pei et al [53]. This indicates that neither second-phase particle pinning effect nor solute segregation drag effect alone can fully account for the occurrence of AGG. We believe that solute segregation of alloying element Y along grain boundaries and its resulting drag effect are a significant factor affecting the grain growth behavior in the current work, however, it was demonstrated in that the levels of grain boundary segregation were independent of grain boundary misorientation angle [57]. Thus any macroscopic anisotropy in GB mobility in these alloys cannot be attributed to differences in GB segregation. However localized fluctuations in GB segregation may play a role [85]. Furthermore, it is important to consider alternative mechanisms. As summarized by Pei et al. [53], the potential contributing factors include: (1) initial size advantages, (2) anisotropic grain boundary characteristics, such as differences in mobility and energy between high- and low-angle boundaries or special misorientation relationships like coincidence site lattice (CSL) boundaries, (3) solid-state wetting effects, (4) gradients in residual dislocation density, and (5) variations in grain boundary complexion and local fluctuations in GB segregation [85]. Several of these mechanisms may be relevant in the current study. Specifically, the broadened $D/\overline{D}$ distributions and the accelerated growth of a subset of grains point to anisotropic boundary mobility effects, while the transient reversion from AGG to NGG suggests dynamic processes such as dislocation density relaxation or complexion transitions.

In the present Mg-Y alloys, a small number of relatively large grains was observed in the initial microstructure (**Figure 1**). However, the overall grain size distribution closely followed a Gaussian distribution. Thus, the initial size advantage alone cannot account for the occurrence of ALGs, particularly since only a limited number of grains with preferential orientations exhibited accelerated growth.

The solid-state wetting mechanism, originally proposed to explain the preferential growth of GOSS-oriented grains in Fe-3%Si alloys, was based on the principle of energy minimization [74–76]. According to this theory, low-energy subgrain boundaries expand at the expense of high-angle boundaries with a much higher energy, facilitated by favorable wetting conditions at triple junctions. In silicon steels, such low-energy subgrain boundaries are often associated with GOSS-oriented grains, which tend to recover rather than fully recrystallize after the deformation due to their lower stored energy [74–76]. However, the initial microstructure of the Mg-Y alloys investigated here was nearly fully recrystallized. Consequently, the possibility of having a considerable fraction of sub-grain boundaries to induce the solid-state wetting for the AGG occurrence was low.

Another potential driver of AGG is the residual dislocation density gradient (DDG). Variations in the stored dislocation energy among adjacent grains after recrystallization can contribute an additional driving force for grain boundary migration during subsequent annealing [86,87]. Grain boundaries adjacent to grains with lower dislocation density tend to migrate more rapidly than those next to grains with higher dislocation density, in an effort to minimize the total free energy of entire system. Pei et al. hypothesized that the driving force induced by DDG provided grains with $(0001)<11\bar{2}0>$ orientation parallel to the rolling direction a growth advantage during the early grain growth stage in pure Mg [53]. To evaluate the role of DDG in the present study, kernel average misorientation (KAM) analysis on EBSD data was performed on Mg-Y alloys to estimate the dislocation density at various grains and grain boundaries. However, no clear correlation between KAM intensity and grain boundary characteristics could be established based on our analysis.

The last mechanism to be considered is grain boundary complexion transition. In this mechanism, AGG found in various doped alumina systems was associated with the presence of grain boundaries with distinct structures and compositions, referred to as complexions [88–91]. The difference in grain boundary mobility, arising from varying degrees of structural disorder associated with different grain boundary complexion types, would promote AGG [88–91]. As discussed earlier, solute segregation at grain boundaries was observed in the current studied Mg-Y alloys. This segregation likely alters the interfacial structure of the grain boundaries, although detailed investigation into grain boundary structure has not yet been conducted. Variations in grain boundary plane and misorientation could lead to the coexistence of multiple complexion types or affect the stability of a given complexion [90]. Consequently, differences in grain boundary mobility may emerge, potentially serving as a factor affecting the observed AGG in this study.

### 4.3.3 Texture evolution during grain growth

Given the texture-weakening effect of rare-earth (RE) alloying elements in Mg alloys, the resulting textures of both RE-free and RE-containing Mg alloys after extrusion vary depending on the extrusion parameters [6,92]. <10$\bar{1}$0> fiber, <2$\bar{1}$$\bar{1}$0> fiber and the <10$\bar{1}$0>-<2$\bar{1}$$\bar{1}$0> double fiber texture have been observed, as grain orientation aligned parallel to the extrusion direction [4–7,92–97]. In the current investigated Mg-Y alloys, a weak <2$\bar{1}$$\bar{1}$0> fiber texture formed after the extrusion procedures, as shown in **Figure 1**. Upon annealing, as grain growth progressed, inverse pole figure texture analysis revealed a shift in the radial texture from <2$\bar{1}$$\bar{1}$0> to <10$\bar{1}$0>. Similar texture evolution, which is associated with preferential grain growth, has been reported in pure Mg and various Mg-alloys during normal and abnormal grain growth processes. For instance, a study on hot-rolled commercial pure Mg suggested that grains with orientations of (0001)<11$\bar{2}$0>//rolling direction exhibited a growth advantage over other basal-oriented grains during AGG [53]. This behavior was hypothesized to arise from gradients in residual dislocation density stored in grains post-recrystallization [53]. A separate investigation on extruded AZ31 Mg alloy sheets revealed that grain growth led to significant texture evolution [79]. In that study, grains with prismatic planes {11$\bar{2}$0} parallel to the sheet surface tend to grow at the expense of grains with other orientations, which were found through normal grain growth in the mid-layer of the sheet and abnormal grain growth in the outer layer of the sheet. Eventually the predominantly basal texture is not observed neither in the mid-layer nor the outer surfaces after the annealing heat treatment [79]. A study on AZ61 sheets showed that grains with tendency to grow abnormally were those with prismatic planes {11$\bar{2}$0} parallel to the sheet plane and <10$\bar{1}$0> directions aligned with the ED [46]. This was rationalized from the perspective of texture effect: in strongly basal-textured microstructures, high-angle grain boundaries (HAGBs) between prismatic and basal grains possess higher energy and mobility than low-angle boundaries between basal grains, making prismatic grains more prone to growth [46]. In an indirect-extruded Mg-1wt.%Gd alloy with a strong textured <2$\bar{1}\bar{1}$1>//ED recrystallized microstructure, abnormal growth of <10$\bar{1}$0>//ED grains occurred after prolonged annealing [98]. This was also attributed to the texture effect, where HAGBs between <2$\bar{1}\bar{1}$1> grains and <10$\bar{1}$0> grains have higher mobility than lower angle boundaries between <2$\bar{1}\bar{1}$1> grains themselves [98].

In the present study, the main grain orientations parallel to the ED in the initial microstructure were (0001)<2$\bar{1}\bar{1}$0> and (0001)<10$\bar{1}$0>, while the former orientation was dominant, as shown in **Figure 1**. EBSD analysis of the grown grains, especially the ALGs, suggested that <10$\bar{1}$0>//ED grains had the advantage to grow, leading to the observed shift in maximum texture intensity from <2$\bar{1}\bar{1}$0> to <10$\bar{1}$0> during grain growth (**Figure 9**). Similar to previous studies [46,98], this texture effect was also believed to contribute to the preferential growth observed in this study, as HAGBs between <10$\bar{1}$0> and <2$\bar{1}\bar{1}$0> grains likely exhibit higher mobility than boundaries between <2$\bar{1}\bar{1}$0> grains. In addition, it is worth noted that the texture transition between <10$\bar{1}$0> and <2$\bar{1}\bar{1}$0> components has been reported as a feature on the recrystallisation behavior of metals with hexagonal structure, since those boundaries having a 30° misorientation around the *c*-axes have a high mobility [51,64].

### 4.4. Solute drag effect in Mg alloys

#### 4.4.1. Description of the CLS solute drag model

The model for solute drag effects on grain boundaries (GBs) proposed by Cahn [65] is employed in this study, which was an extension of Lücke and Stüwe's original model [99], and is often referred to as the Cahn-Lücke-Stüwe (CLS) model. This model assumes that a solute (or impurity) atmosphere at a grain boundary exerts a drag force on a migrating GB and slows it down. The CLS formulation estimates this drag force by solving for an evolving solute concentration profile at a GB moving with a particular velocity. The main equation for the CLS model is written in terms of a total drag pressure, *P*, required to move a GB at velocity $V$, expressed as a sum of the intrinsic drag for a pristine GB, $P_0(V)$, and the drag force exerted by the solute/impurities, $P_i$:

$$P = P_0(V) + P_i \qquad (3)$$

$$P = P_0(V) - N \int_{-\infty}^{+\infty} (c(x) - c^{bulk}) \frac{d\Delta F^{seg}}{dx} dx \qquad (4)$$

Where, $c(x)$ – is the solute concentration profile around the GB, $c^{bulk}$ – is the bulk solute concentration, $\Delta F^{seg}$ – is the solute segregation free energy, and $N$ – is the number of solutes at the GB. The solute concentration profile $c(x)$ also depends on the solute diffusivity, $D(x)$, through a flux relationship detailed in Reference [65].

A linear profile is assumed for the solute segregation free energy, $\Delta F^{seg}(x)$ in a $\delta = 1$ nm region surrounding the GB, wherein their respective magnitudes are maximum at the GB plane ($x = 0$). In the CLS model [65], using the assumed profiles and approximations drawn from extreme cases of GB velocity, the following equation is derived for the relationship between drag pressure and velocity:

$$P = \lambda V + \frac{\alpha c^{bulk} V}{1+\beta^2 V^2}. \qquad (5)$$

Here, $\alpha = \frac{N\delta(k_B T)^2}{\Delta F_0^{seg} D} \left( \sinh \frac{\Delta F_0^{seg}}{k_B T} - \frac{\Delta F_0^{seg}}{k_B T} \right)$, $\beta^2 = \frac{\alpha k_B T \delta}{2N\Delta F_0^{seg^2} D}$, and, $\lambda$ – is the intrinsic drag coefficient that is the reciprocal of the intrinsic mobility of a pure GB (Note: $\Delta F_0^{seg} = \Delta F^{seg}(x = 0)$). Equation (5) will be used in all further calculations of the solute drag effect, and the parameters used within are summarized in Table 2. Values for bulk solute diffusivities at different temperatures for the two solutes are obtained using the diffusion constant and activation energy barriers computed using first-principles simulations in Reference [100]. For simplicity, the solute diffusivity is assumed to be constant around the GB region. The solute drag calculations in this study are focused on the temperature range, $T = 227 - 527°C$, relevant to experiments and the thermomechanical processing of Mg alloys. In the following sections, methods used to estimate segregation free energy for Mg-Y and Mg-Al systems are described in detail.

Table 2. Parameters used in the CLS model as given by Equation (5).

| | |
|---|---|
| Y diffusivity [100] | $D_0^Y = 1.41 \times 10^{-6} \frac{m^2}{s}$, $Q^Y = 1.198$ eV |
| Al diffusivity [100] | $D_0^{Al} = 1.8 \times 10^{-6} \frac{m^2}{s}$, $Q^{Al} = 1.31$ eV |
| GB extent, $\delta$ | 1 nm |
| Solute atoms at GB, $N$ | 1 nm$^3$ |
| Intrinsic GB drag, $\lambda$ | $10^4$ GPa.s/m |

### 4.4.2. Finite temperature segregation energetics for Mg alloys:

**Mg-Y system:** A key input for the CLS model is the solute segregation free energy, $\Delta F^{seg}$. In a previous study [101], the spectral segregation approach [102] was applied in conjunction with free energy calculations of Y segregation using an MEAM potential [103] at Mg symmetric tilt GBs to construct an experimentally validated predictive model for finite temperature grain boundary segregation. The segregation free energy calculations at GB sites, $\Delta F_i^{seg}$, were performed using thermodynamic integration methods [104] implemented in LAMMPS [105], which provide a complete description of finite temperature internal energy and vibration entropy. A stacking cross-validation machine learning model with physics-informed descriptors was then used to learn computed $\Delta F_i^{seg}(T)$ for less than 100 GB sites and predict free energy distributions for thousands of sites over a temperature range $27 - 527°C$. With the predicted $\Delta F_i^{seg}(T)$ distributions, estimates were made for average Y segregation tendency at finite temperatures, and it was found that equilibrium segregation concentration at high temperatures (T = $227 - 527°C$) are much lower than those estimated using segregation enthalpy, $\Delta E_i^{seg}$. These free energy-based segregation estimates also agree well with experimental observations of segregation, highlighting an important aspect that slow-diffusing solutes like Y can only achieve equilibrium segregation states corresponding to high temperatures relevant to the thermomechanical processing of Mg alloys. Detailed comparison of the simulation results for Y segregation and experimental observations was presented in Reference [57]. The effective $\Delta F_Y^{seg}(T)$ values from our study are shown in Figure 15(a) for different temperatures and bulk Y concentrations, $c_Y^{bulk} = 0 - 13$ wt.%, which will be used in the following sections. The previous studies indicate that Y shows attractive segregation tendency over the temperature and bulk concentration ranges considered, and the corresponding average GB Y concentration, $c_Y^{gb}$, are shown in Figure 15(b).

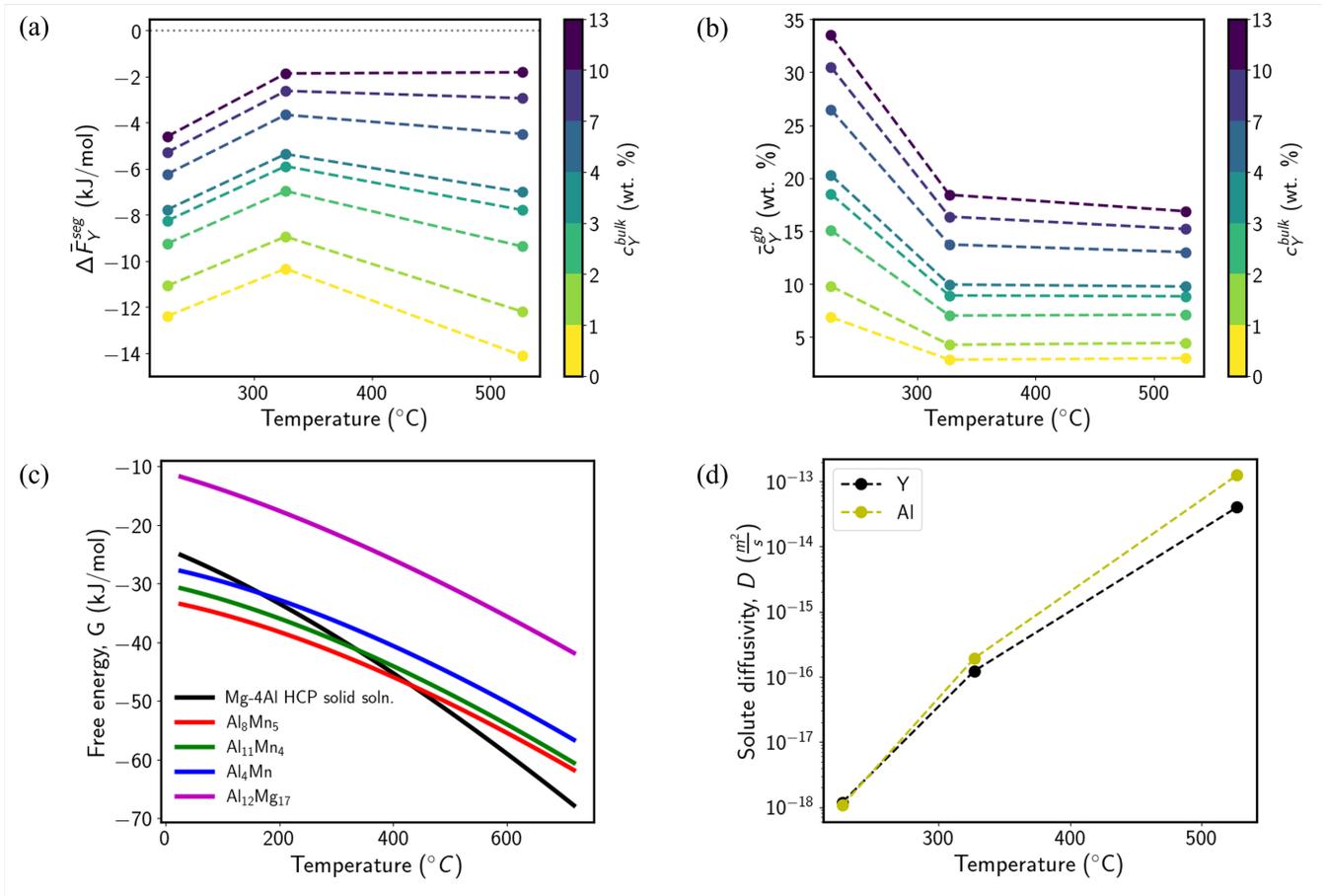

**Figure 15. (a) Effective Y segregation free energy ($\Delta \bar{F}_Y^{seg}$) versus temperature for a range of bulk Y concentrations, $c_Y^{bulk}$; (b) GB segregation concentration for Y versus temperature for a range of bulk Y concentrations, $c_Y^{bulk}$, corresponding to $\Delta \bar{F}_Y^{seg}$ values in (a); (c) CALPHAD-based bulk free energies versus temperature for Al-based phases and Mg-4Al solid solution; (d) solute diffusivities for Al and Y versus temperature determined using Ref. [98].**

**Mg-Al system:** Mg-Al alloys have been studied extensively experimentally and as mentioned in previous sections, reported observations of phase precipitations in these alloys [3,51] far exceed reports of GB solute segregation. One study by Pei et al. [56] shows that high temperature processing can dissolve intermetallic phases in Mg-Al-Ca-Zn alloys and make solutes available for co-segregation. Therefore, the development of models for Al segregation at Mg GBs is tricky, requiring a consideration of the precipitation of bulk or GB phases, that can alter the concentration of solute atoms in solid solution that can segregate to GBs. Additionally, it is likely that the solute effect on GB migration in Mg-Al alloys possibly has distinct regimes, one involving Zener pinning phenomena due to second-phase particles, and the other involving the CLS model description of solute drag.

In the current study, a phase dissolution temperature, $T_{sol}$, is estimated for Mg-Al using bulk CALPHAD-based free energy calculations [106,107] to compare the thermodynamic stability of Al-based phases and Mg-4Al in solid solution. Following established phase diagrams, the $\gamma-Mg_{17}Al_{12}$ phase and several possible Al-Mn phases ($Al_8Mn_5$, $Al_{11}Mn_4$, and $Al_4Mn$) are considered in our calculations; the latter are included following the assumption that Mg-Al alloys usually consider trace Mn quantities that form phases with Al [3,51,108]. The Redlich-Kister formulation [109] is used to compute free energies of the pure elements and solid solution, whereas for the phase energies the sublattice-based compound energy formalism is used [107]. All parameters for the free energy calculations are obtained from References [110–112]. While the results from the CALPHAD free energy calculations in Figure 15(c) are not exact, they provide an estimate of the temperature ($T_{sol} \sim 430°C$) at which an Mg-4Al alloy in solid solution becomes more thermodynamically favorable compared to the secondary phases. The estimated temperature is conservative compared to solution heat treatment temperatures ($\sim 530°C$) reported in experiment for Mg-Al multicomponent alloys [56]. The CLS model will only be applied to Mg-Al at temperatures higher than $T_{sol}$ where the solid solution is stable, which will be discussed further in the following section.

### 4.4.3. Results

**Mg-Y system:** Results from CLS model calculations for the Mg-Y system are presented first as they are illustrative of the influence of solute segregation thermodynamics versus solute diffusion kinetics on GB solute drag. Considering a range of velocities, $V = 10^{-4} - 10^4 \frac{\mu m}{s}$, relevant to recrystallization and grain growth, the total drag pressure, $P$, is computed for each temperature and Y bulk concentration, with key findings shown in Figure 16. The plotted drag pressure, $P$, is normalized by the quantity $P^{max}_{13Y,227°C}$, which is the maximum drag pressure in our calculations across all temperatures and bulk Y concentrations, occurring at a bulk Y concentration, $c_Y^{bulk} = 13$ wt.%, and a temperature, $T = 227°C$. These results are obtained using the segregation free energies from Figure 15(a) and solute diffusivities from Figure 15(d) within the CLS framework given in Sec. 4.4.1.

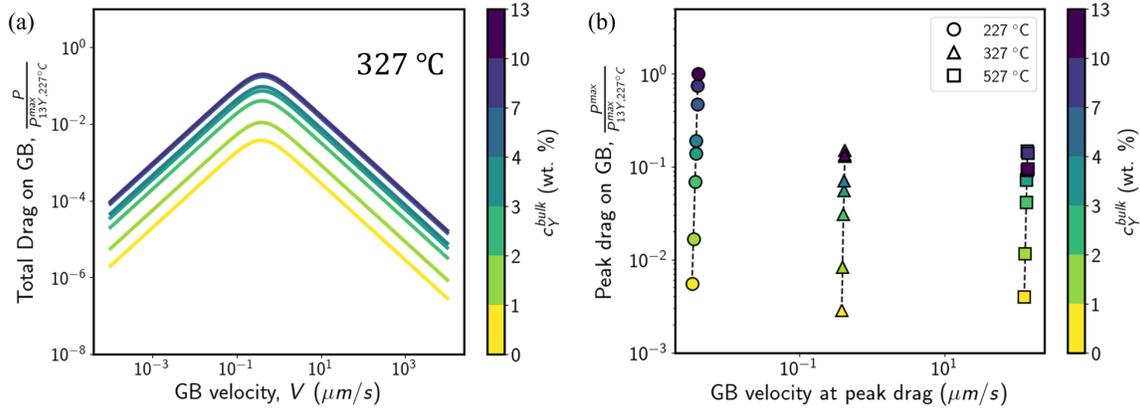

**Figure 16. (a) Total drag pressure, $P$, versus GB velocity, $V$, for a range of bulk Y concentration, $c_Y^{bulk}$, at temperature $T = 327$ °C; (b) peak solute drag pressure, $P_{max}$. versus corresponding GB velocity for a range of bulk Y concentration, $c_Y^{bulk}$, at temperatures, $T = 227, 327,$ and $527$ °C. The drag pressure, $P$, in the above figures is normalized by the maximum drag pressure across all our calculations, $P^{max}_{13Y,227°C}$, which occurs for Mg-13Y at $227°C$.**

In Figure 16 (a), the total drag pressure, $P$, versus GB velocity, $V$, is shown for a $c_Y^{bulk}$ range from 0.1-13 wt. % at temperature $T = 327°C$. The plot shows two branches that make up the essence of the CLS solute drag model. The initial $\log(P) - \log(V)$ linear dependence represents a dragged regime, where GB velocity is low and the segregated Y solute atmosphere moves along the GB. The point where the drag pressure peaks is the point of breakaway, where the GB will leave the segregated solute atmosphere. At velocities higher than breakaway, the driving force required to move the GB are lower since the solutes do not diffuse fast enough to catch up with the grain boundary leading to a lowered solute drag pressure.

Following similar calculations for temperatures $T = 227$ and $527\,°C$, the peak drag pressure, $P_{max}$, and corresponding breakaway GB velocity are recorded for each $c_Y^{bulk}$ and plotted in Figure 16 (b). Both Figures 16 (a) and (b), show a correlation with between the drag pressure, $P$, and the segregated Y concentration, $c_Y^{GB}$. As noted before, the maximum drag pressure is found to be at the strongest segregation state at $T = 227°C$ for $c_Y^{bulk} = 13$ wt. %, where the solute enrichment at the GB is almost 3 times the bulk concentration. Additionally, Figure 16(b) shows that the breakaway GB velocity corresponding to $P_{max}$ increases with temperature but is mostly constant for changing $c_Y^{bulk/gb}$ at a particular temperature.

The first phenomenon of the $P - c_Y^{gb}$ correlation is a consequence of equilibrium segregation thermodynamics, and is intuitive to understand, since a GB with a higher segregated solute concentration should experience a higher drag pressure at the same GB velocity and temperature. Previous studies on the solute drag effects in Mg alloys used a single effective segregation energy for solutes at all temperatures [15, 45]. The more rigorous segregation free energy calculations used in the present study provide new insights like the fact that the $c_Y^{gb}$ gets saturated at higher temperatures ($T > 300\,°C$) for a given $c_Y^{bulk}$. Therefore, at higher temperatures, there is a corresponding saturation in the maximum possible solute drag pressure, $P_{max}$. This is evident in Figure 16(b), where the $P_{max}$ increases monotonically with $c_Y^{bulk}$ at the lowest temperature, $T = 227\,°C$, while there is a distinct saturation of $P_{max}$ at higher $c_Y^{bulk}$ for the highest temperature, $T = 527\,°C$.

The second phenomenon of higher breakaway GB velocity at higher temperature is a result of increasing solute diffusivity as shown in Figure 15(d). The implication here is that faster-diffusing solutes can keep up with a moving GB, maintaining a solute drag regime at higher GB velocities before breakaway is achieved. A less intuitive consequence of this is that a lower GB segregation state and a slower diffusivity at lower temperatures can have a more effective solute drag phenomenon for a given GB velocity. This idea is similar to the discussion by Robson et al. [15,45], where GB velocities are assumed to be $\sim 10^{-4} - 1 \frac{\mu m}{s}$, and it is concluded that solutes that show peak drag forces comparable to the recrystallization/grain growth driving force ($\sim 1 - 10$ MPa) are more effective at impeding GB migration kinetics.

Overall, considering these two phenomena it can be asserted that Y is more effective at slowing down GB migration kinetics at the lower end of temperatures at which thermomechanical processing is carried out, i.e. $T \sim 200 - 400°C$, where there is combined effect of a strong segregation state and intermediate solute diffusivity creates a solute drag pressure comparable to recrystallization/grain growth driving forces.

These findings from the CLS model can explain several experimental observations presented in previous sections. Considering firstly recrystallization kinetics in Figure 4, which shows the percentage of recrystallized grains versus annealing time for Mg-1Y and Mg-7Y for two temperatures. The annealing times at which 50% recrystallization is achieved are 7-8 times higher at 350°C compared to 400°C for each bulk Y concentrations. This result is easily explained by the higher solute segregation and hence, a higher solute drag effect, at lower temperatures for each bulk concentration as predicted by the combined finite temperature segregation energetics and the CLS model in this study. A more surprising aspect of the experimental results is that the Mg-7Y at 400°C (higher $c_Y^{gb}$, faster $D_Y$) achieves 50% recrystallization four times faster than Mg-1Y at 350°C (lower $c_Y^{gb}$, slower $D_Y$). This supports the previous assertion that it not just a strong segregation state or faster solute diffusivity that result in a strong/weak solute drag effect, but rather a combined effect at the lower range of alloy processing temperatures where a solute drag comparable to the GB migration driving forces is generated.

Similar arguments can assist in understanding experimental results for grain growth behavior, for instance Figure 6, which shows grain diameter versus annealing time for Mg-7Y at a range of temperatures, $T = 400 - 560°C$. The grain diameter does not show significant variation for long annealing times for temperatures lower than 500°C, and at higher temperatures the onset of grain growth occurs at smaller annealing times. The modeling results indicate the segregation state for Mg-7Y will get weaker and the solute diffusivity will increase as temperature increases. The assertion that Y solute drag is more effective at the lower range of processing temperatures holds in this case as well, where the drag pressure magnitudes are comparable to the driving force for grain growth.

While the CLS model with accurate segregation free energetics is demonstrated to be useful for understanding general recrystallization and grain growth phenomena in Mg-Y, there are some limitations in understanding certain nuances. For instance, it is presently difficult to explain the increasing spread in measured grain diameters with increasing temperature and annealing time in Figure 6, especially for temperatures higher than 500°C. This is analyzed in experiment by measuring a parameter for abnormal grain growth (AGG) that is shown in Figure 14. Such AGG phenomena indicates that grain growth in alloys involves some stochastic processes, where certain GBs breakaway from their solute atmosphere. These processes can be governed by complex energy landscapes at the solute segregated GB, solute diffusivity, and anisotropy in intrinsic GB mobility, that are not presently included in the CLS model formulation. Recent theoretical kinetic Monte Carlo models for solute drag proposed by Mishin formulate GB migration as a random walk combined with solute pinning events and GB roughening transitions can help overcome these limitations [113,114]. In future studies, connecting the parameters of these theoretical models to real alloys could provide a more comprehensive understanding of the effects of solute segregation and diffusion effects on various GB migration phenomena including AGG.

**Comparing solute drag effects for Mg-Y and Mg-Al systems:**

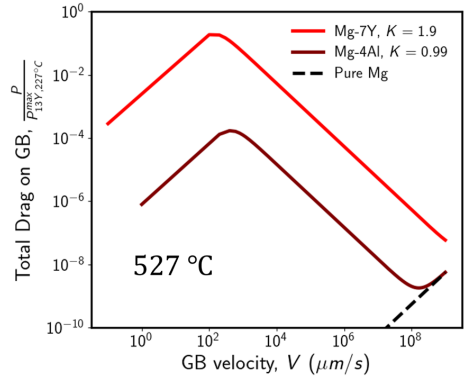

**Figure 17. Total drag pressure versus GB velocity at temperature, $T = 527°C$, for a Mg-7Y alloy (using $\Delta F^{seg}$ computations in Figure 15(a)), a Mg-4Al alloy and pure Mg. The drag pressure in this figure is normalized by the maximum drag pressure for Y calculations, $P^{max}_{13Y,227°C}$, which occurs for Mg-13Y at $227°C$.**

In the previous discussion, it was shown that the CLS model is successful in understanding solute drag effects on general experimental observations of recrystallization and grain growth behavior in Mg-Y alloys. Here, we apply the CLS model to explain experimental results from Figure 13 that compare grain growth behavior in pure Mg, Mg-4Al and Mg-7Y at different temperatures.

Using CALPHAD-based bulk thermodynamics of Mg-Al systems detailed in Sec. 4.4.2, a temperature ($T_{sol} \sim 430°C$) was estimated at which the Mg-4Al solid solution becomes more stable than the secondary phases for Mg-Al and Al-Mn (see Figure 15(c)). Below or close to $T_{sol}$, GB migration in Mg-Al is more likely to be impeded by Zener pinning mechanisms due to the presence of secondary phases, and it is hypothesized that the experimental results for Mg-4Al in Figure 13(a) at 435°C are due to this effect. The alternative explanation using the CLS model for the similarity in the recrystallization kinetics for Mg-7Y and Mg-4Al in Figure 13(a), would be that Al has achieved the same $c^{gb}$ as Y at that temperature, requiring a highly attractive segregation tendency for Al. Since there is little to no evidence in existing literature that Al solutes can segregate as strongly as Y without the presence of co-segregating elements, Zener pinning is the most probable mechanism behind Mg-Al grain growth kinetics seen in Figure 13(a).

For the experimental observations in Figure 13(b), since the temperature of interest is well above $T_{\text{sol}}$, the CLS model is applied at a similar temperature, $T = 527°C$, and the drag pressures for pure Mg, Mg-4Al, and Mg-7Y are compared. For pure Mg, the only source of drag is the intrinsic GB drag according to a linear relationship (see Equation. (5)). The Mg-7Y calculations follow the same workflow as the results in the previous section. For Mg-4Al, since the segregation state was not known, several different values of enrichment ratio, $K = \frac{c_{\text{Al}}^{\text{gb}}}{c_{\text{Al}}^{\text{bulk}}}$, were tested in the range $0.5 - 1.5$. It was found that in the slightly desegregating to slightly segregating range ($\sim 0.9 - 1.1$) range, the solute drag trends were very similar and the present results assume a $K_{\text{Al}} \sim 0.99$.

Figure 17 shows the results for drag pressures in pure Mg, Mg-4Al, and Mg-7Y versus GB velocity at 527°C. The solute drag pressure for Mg-7Y is over three orders of magnitude higher than Mg-4Al, which in turn is several orders of magnitude higher than a pure Mg GB, up to the point where the GB velocities are high enough that alloyed GBs approach intrinsic drag behavior. These drag pressure trends compare well with experimental observations in Figure 13(b). Pure Mg with the lowest estimated drag pressure, shows the fastest rate of increase in grain diameter, followed by Mg-4Al, whereas Mg-7Y shows little change in grain diameters over long annealing times.

To summarize, in this section the strengths and weaknesses of the CLS solute drag model are demonstrated in understanding experimental recrystallization and grain growth phenomena in Mg alloys. For the Mg-Y system, using an accurate model for segregation free energetics with the CLS model provides useful insights that Y solute drag is most effective at lower thermomechanical temperature ranges where a combination of a strong segregation state and intermediate solute diffusivity are comparable to driving forces governing recrystallization/grain growth. The CLS model was also used to compare solute drag effects in Mg-Y and Mg-Al that primarily differ in their segregation tendency. Some gaps and future directions were identified in the CLS implementation that can improve the understanding of underlying phenomena like abnormal grain growth in microstructure evolution during thermomechanical alloy processing.

## 5. Conclusions

This study investigated the grain structure evolution in two Mg-Y alloys during annealing heat treatment with and without prior deformation. The role of Y in the recrystallization and grain growth behavior was characterized. The main findings are:

1. A comparison between grain evolution in Mg-1Y and Mg-7Y alloys revealed that higher Y additions significantly slowed both recrystallization and grain growth processes. In both alloys, grain structure evolution exhibited strong temperature dependence, with higher annealing temperatures accelerating the kinetics.

2. Static recrystallization in Mg-Y alloys followed a two-stage process: (1) an initial rapid stage characterized by JMAK exponents of ~3 and (2) a retarded stage with JMAK exponents one to two orders lower. This behavior was attributed to heterogeneous nucleation of recrystallized grains.

3. Compared with pure Mg and Mg-4Al, Mg-1Y and Mg-7Y alloys required substantially longer annealing times to achieve equivalent levels of recrystallization. This sluggish kinetics was associated with the possible retardation of diffusion process by Y addition and, more importantly, to solute drag from Y segregation at grain boundaries, which reduced boundary mobility.

4. Grain growth in Mg-Y alloys was slower than in pure Mg and Mg-4Al alloy of similar processing history. The retardation was attributed to solute drag from Y segregation, which enhanced microstructural thermal stability.

5. In Mg-7Y alloy, the grain growth initially proceeded through the abnormal grain growth mode, but this mode was transient and eventually transitioned into normal grain growth.

6. Different possible mechanisms for the occurrence of abnormal grain growth were discussed. The anisotropic grain boundary characteristics (particularly the mobility difference of high and low angle boundaries) combined with potential grain boundary complexion formation induced by Y segregation were believed to be main drivers for the abnormal grain growth in this system.

7. Preferential growth of grains with $(0001)<10\bar{1}0>$ orientation parallel to the extrusion direction was observed. This was rationalized by the texture effect, since the high angle boundaries between $<10\bar{1}0>$ grains and $<2\bar{1}\bar{1}0>$ grains exhibited higher mobility than the low-angle boundaries between $<2\bar{1}\bar{1}0>$ grains.

8. The utility of the CLS solute drag model in understanding experimental observations of solute effects on GB migration kinetics was demonstrated for different Mg alloys. For the Mg-Y system, using an accurate model for finite temperature segregation free energetics with the CLS model provides new insights on the sensitivity of solute drag pressures to equilibrium segregation states and solute diffusivity at different temperatures. It is found that Y solute drag is strongest at the lower range of thermomechanical processing temperatures where the segregation state is strong and solute diffusivity is intermediate. The CLS model is also used to compare solute drag effects in Mg-Y and Mg-Al at high temperatures where the primary difference lies in the segregation states.

Overall, the results highlight the critical role of solute segregation, particularly that of Y in controlling the recrystallization and grain growth behavior of Mg-Y alloys. These insights provide valuable guidance for the design of thermally stable magnesium alloys with refined and stable microstructures.

**Acknowledgement**

This work was supported by PRISMS (PRedictive Integrated Structural Materials Science) center which is located at University of Michigan and funded by U.S. Department of Energy, Office of Basic Energy Science, Division of Materials Science and Engineering (Grant award number DE-SC0008637). We acknowledge the access to and the support from Michigan Center for Materials Characterization (MC2) at University of Michigan and CanmetMATERIALS, Natural Resources Canada who provided the materials used in this investigation. We also gratefully acknowledge Alex Pielack and Ryan Gast who assisted in heat treatments and sample preparation.

**Data Availability**

The experimental data supporting this study is available on the Materials Commons at xxx (DOI to be established upon article acceptance).

**Conflict of Interest**

The authors declare that they have no known conflict of interest.